\documentclass[fleqn,usenatbib]{mnras}

\usepackage{newtxtext,newtxmath}
\usepackage{mathptmx}

\usepackage[T1]{fontenc}

\usepackage{graphicx}	
\usepackage{amsmath}	
\usepackage{amssymb}	

\title[Finding galaxy groups via radio surveys]{A new method for finding and characterizing galaxy groups via
  low-frequency radio surveys} \author[J. H. Croston]
{J. H. Croston$^{1}$\thanks{Email: Judith.Croston@open.ac.uk}, 
  J. Ineson$^{2}$, M. J. Hardcastle$^{3}$ and B. Mingo$^{1}$\\$^1$ School of Physical Sciences, The Open
  University, Walton Hall, Milton Keynes, MK6 7AA, UK\\$^2$ School of Physics and Astronomy,
University of Southampton, Highfield, Southampton SO17 1BJ, UK\\$^3$
Centre for Astrophysics Research, School of Physics, Astronomy and
Mathematics, University of Hertfordshire, Hatfield, Herts, AL10 9AB, UK.}

\date{Accepted XXX. Received YYY; in original form ZZZ}
\pagerange{\pageref{firstpage}--\pageref{lastpage}}
\pubyear{2017}

\defcitealias{ineson17}{I17}
\defcitealias{arnaud10}{A10}

\begin{document}
\label{firstpage}
\pagerange{\pageref{firstpage}--\pageref{lastpage}}
\maketitle

\begin{abstract}
  We describe a new method for identifying and
  characterizing the thermodynamic state of large samples of evolved galaxy groups at high redshifts using high-resolution, low-frequency radio surveys, such as those that will be carried out with LOFAR and the Square Kilometre Array (SKA). We identify a sub-population of morphologically regular powerful (FRII) radio galaxies and demonstrate that, for this sub-population, the internal pressure of the radio lobes is a reliable tracer of the {\it external} intragroup/intracluster medium (ICM) pressure, and that the assumption of a universal pressure profile for relaxed groups enables the total mass and X-ray luminosity to be estimated. Using a sample of well-studied FRII radio galaxies, we demonstrate that our method enables the estimation of group/cluster X-ray luminosities over three orders of magnitude in luminosity to within a factor of $\sim 2$ from low-frequency radio properties alone. Our method could provide a powerful new tool for building samples of thousands of evolved galaxy groups at $z>1$ and characterizing their ICM.
\end{abstract}


\section{Introduction}
\label{sec:intro}
The thermodynamic evolution of baryons in forming and evolving galaxy
groups and clusters at high redshift is not currently well
constrained, but is an essential ingredient in cosmological
studies of large-scale structure evolution (e.g. cluster counts,
Sunyaev-Zeldovich (SZ) studies, cosmic shear). With current X-ray
telescopes it is just possible, with enormous expenditure of observing
time, to identify a small number of evolved groups at $z>1$
\citep[e.g.][]{tanaka13,finoguenov15}. SZ surveys are not currently sensitive to masses below $\sim 10^{14}$ M$_{\sun}$. Systematic studies of the evolution of the ICM above
$z=1$ must therefore await next-generation instruments. New, reliable and efficient
methods for identifying and characterizing evolved galaxy groups at
high redshift would therefore be extremely valuable.

One of the major gaps in our understanding of ICM evolution is the
cumulative contribution of AGN feedback from the protogroup/cluster
stage to the present. The lack of observational constraints either on
ICM properties or AGN feedback behaviour at high redshift is a
limitation for cosmological simulations \citep[e.g.][]{mccarthy17,haider16}, and the relative contributions of quasar winds and jets are not well constrained. At present jet population demographics are poorly known
beyond $z \sim 1$, but this situation is rapidly changing with first
results emerging from the LOFAR extragalactic surveys,
revealing the population of low/intermediate luminosity jets at high
redshift \citep[e.g.][]{hardcastle16,williams16,shimwell17}. These new radio surveys have enormous
potential to improve our understanding of AGN feedback, but their interpretation
requires (1) a firm understanding of the physics and energetics of the
diversity of jet sub-populations, so that radio properties can be
translated into energetic impact, and (2) the ability to map jet
populations to the dark matter haloes in which they live.

We have recently completed a comprehensive survey of the environments
of radio-loud AGN, across three orders of magnitude in radio
luminosity and environmental richness \citep{ineson13,ineson15,ineson17}. A key outcome of this work is the demonstration that both
morphology (FRI/II class) and accretion mode (low/high excitation)
influence the locations and impact of jet feedback. We have found a
relationship between radio luminosity and cluster richness for
low-excitation radio galaxies (LERGs), which holds across four orders of
magnitude, with evidence that morphology is a source of scatter in
this relation. No similar relationship was found for high-excitation
radio galaxies (HERGs), which are found almost exclusively to occupy galaxy
groups (at least in the redshift range of $z<0.6$ considered in these studies). Our work suggests that radio galaxies will be excellent indicators of
galaxy group (rather than rich cluster) environments at high redshift, but the LERG $L_{\rm R}$--$L_{\rm X}$
relation we have found \citep{ineson15} (where $L_{\rm R}$ is low-frequency radio luminosity and $L_{\rm X}$ is bolometric X-ray luminosity) has large scatter. In addition, the radio-loud AGN population is likely to become increasingly
dominated by high-excitation sources at high redshift \citep[e.g.][]{best14,williams15,williams16}. Radio luminosity alone therefore does not provide a reliable indicator of a galaxy group environment or allow the prediction of halo mass or X-ray luminosity to within an order of magnitude.

The aim of this paper is to present a new method of identifying and characterizing evolved galaxy group environments, based on
considering the AGN sub-population whose internal physical conditions are best understood. We have shown \citep{croston05,ineson17} via X-ray inverse-Compton (IC) observations of radio lobes,
that FRII radio galaxies have typical magnetic field strengths ($B$) a
factor of a few less than the equipartition value ($B_{\rm eq}$, e.g. \citealt{burbidge56}) calculated with the assumption of no protons. The FRII radio galaxies in the representative sample of \citet{ineson15} have well-constrained
internal pressures from X-ray IC measurements and external pressures
from the ICM X-ray emission \citep[][hereafter \citetalias{ineson17}]{ineson17} allowing us to examine the source dynamics and establish that FRII lobes do not contain a significant proton
population. This result is in contrast to the inferred internal
conditions in low-power (FRI) radio galaxies, where it is well
established that significant proton content and/or departure from
equipartition is required for pressure balance \citep[e.g.][]{croston03,croston08,birzan08,croston14}. Our FRII study has led us to postulate that -- {\it for the FRII jet sub-population only} -- the radio-measured internal pressure should provide a reliable estimate of the {\it external} ICM pressure at the lobe midpoint, where we have found that the lobes are close to pressure balance with the environment (consistent with hydrodynamical simulations of FRII lobe evolution, e.g. \citealt{hardcastle14}).

Below we describe a method to exploit these results, and demonstrate
that FRII radio galaxy samples from wide-area low-frequency radio
surveys (including both low- and high-excitation objects) can be used to find and characterize evolved galaxy groups, potentially enabling the compilation of large catalogues of groups at high redshift.

\section{A method for predicting group-scale environments}
\label{sec:method}
Our aim is to demonstrate that group/cluster X-ray luminosity can be
predicted from radio properties alone. Specifically, for regular FRII
radio galaxies the relatively narrow distribution of both
$B_{\rm obs}$/$B_{\rm eq}$ and midpoint $P_{\rm ext}$/$P_{\rm int}$ \citepalias{ineson17} leads
us to expect that the radio-estimated internal pressure should be a reliable
predictor of external ICM pressure, which can then be used to estimate
the overall environmental richness. In order to have a robust method
that can be applied to large-area surveys, we need to be able to use
easily measured radio properties to identify the subpopulation of
radio-loud AGN for which such predictions are likely to be
reliable. Below we discuss how to select such a sample, before
developing and testing a radio-based method for predicting ICM
luminosity using the sample of \citetalias{ineson17}.

%
\begin{figure*}
\centering{\hbox{
\includegraphics[height=4.8cm]{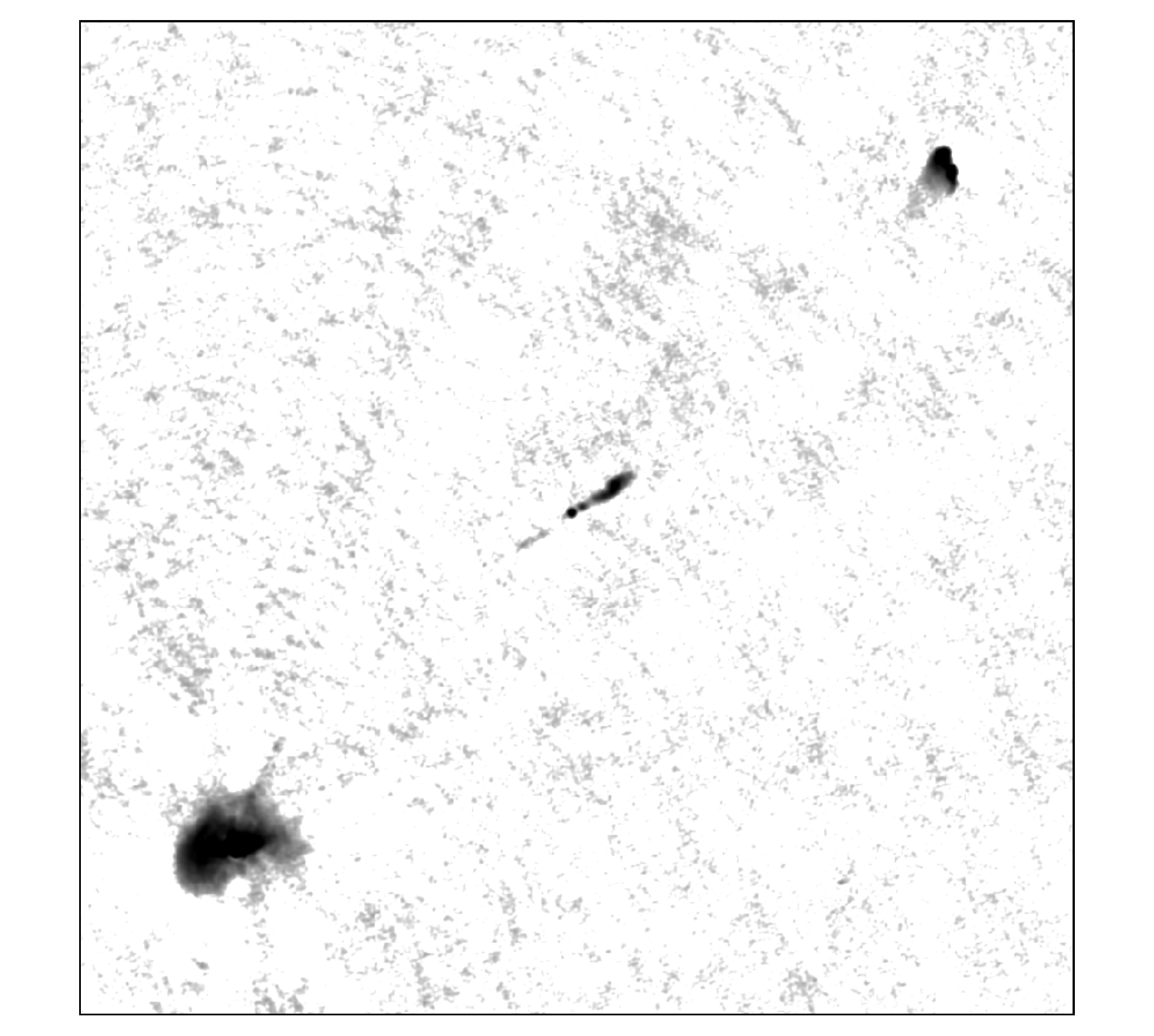}
\includegraphics[height=4.8cm]{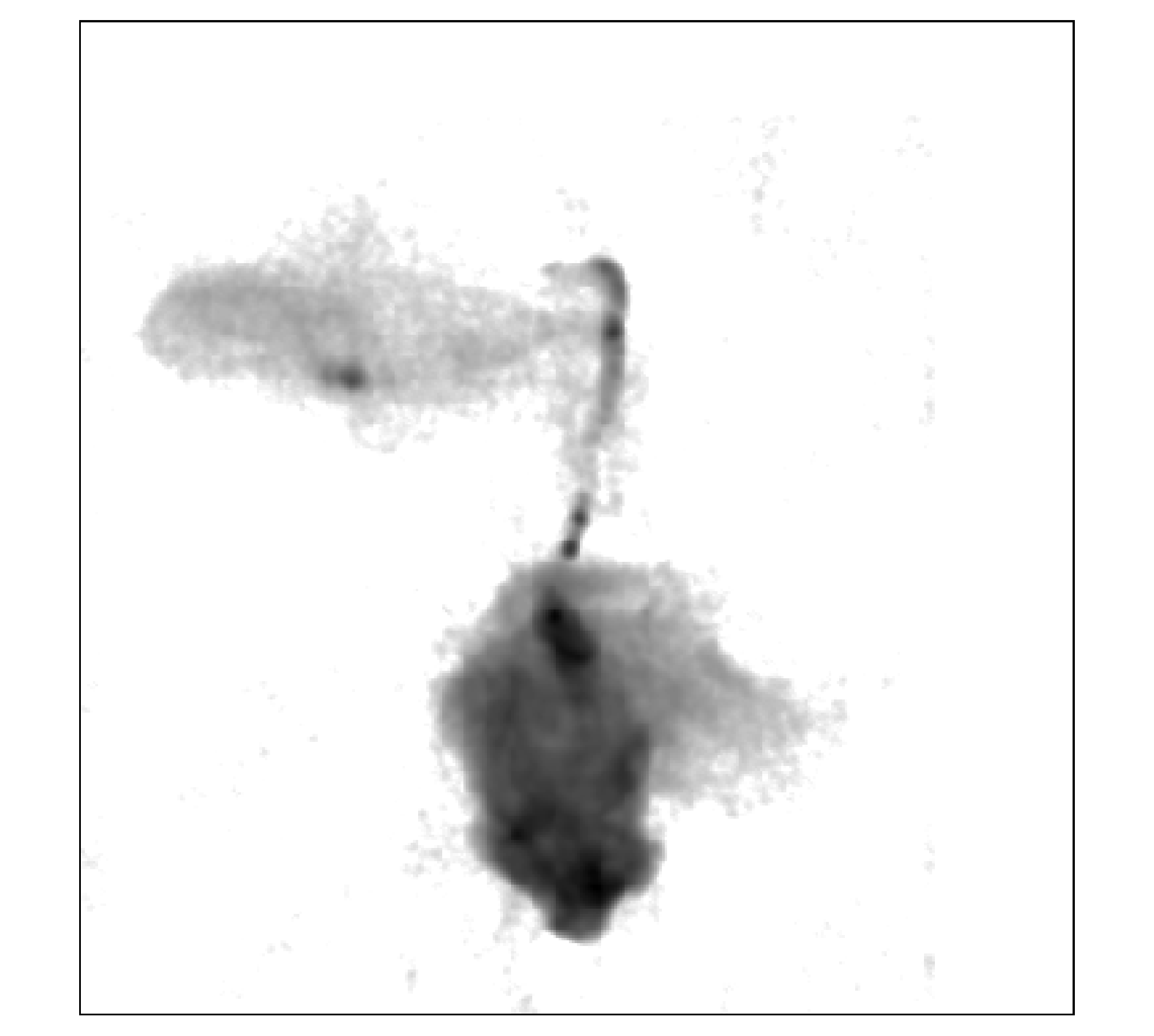}
\includegraphics[height=4.8cm]{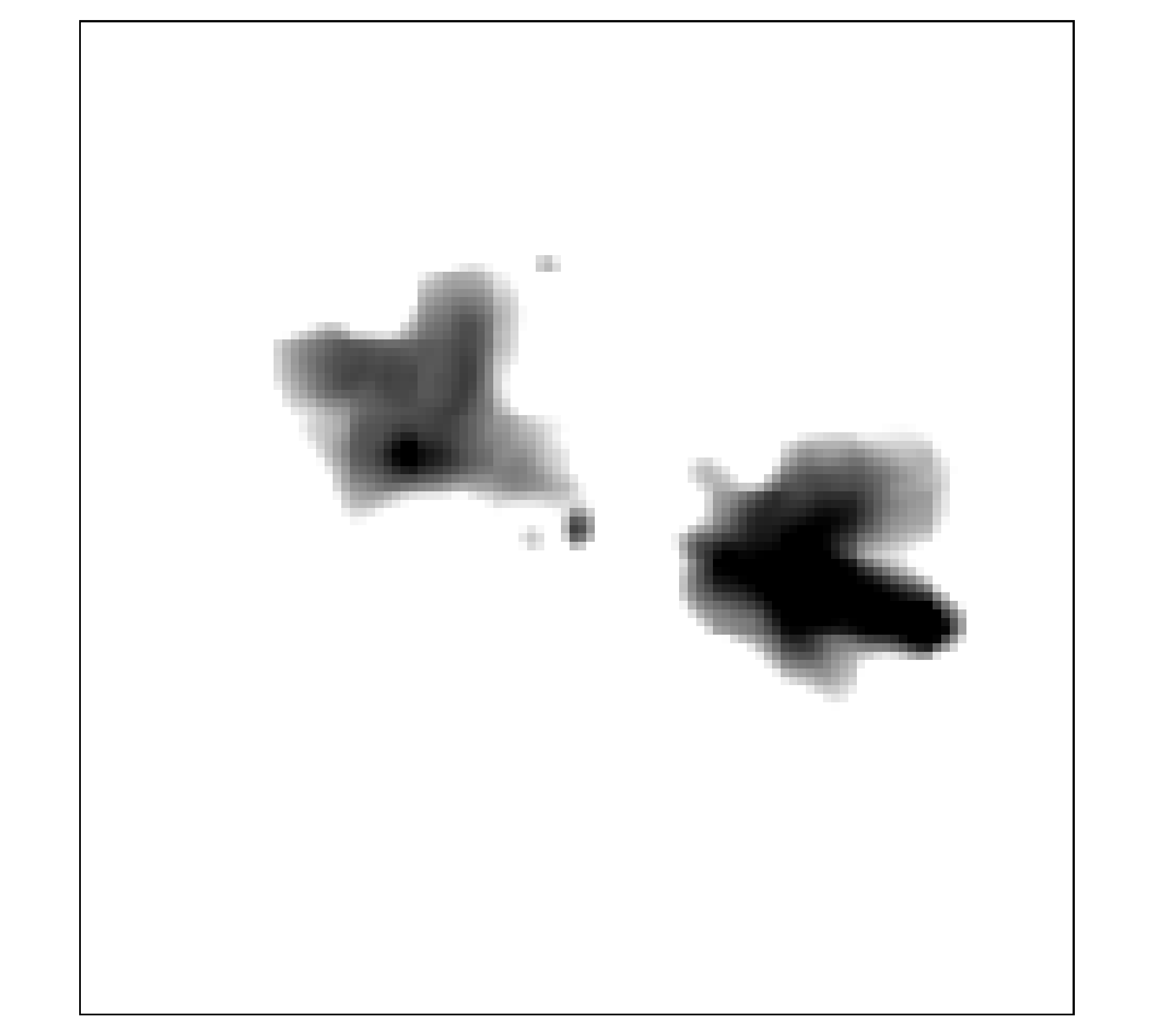}
}}
\caption{Examples of sources excluded from the sample due to irregular morphology or lack of extended lobe emission at low frequencies, from left: 3C\,321, 3C\,433 and 4C\,73.08. Images were obtained from the Atlas of DRAGNs (Leahy, Bridle \& Str\"{o}m: \url{http://www.jb.man.ac.uk/atlas/}).}
\label{fig:exclude}
\end{figure*}
\subsection{Morphological selection}
\label{sec:select}
The sub-sample of radio galaxies whose large-scale environments should
be predictable from radio-lobe pressure estimates must meet the
following requirements:
\begin{itemize}
\item FRII (edge-brightened) morphology, so that proton content is
  expected to be minimal \citep{croston05,ineson17}
\item symmetric ``classical double'' structure, which we postulate
  should be indicative of a relaxed environment with the radio galaxy
  located at the centre of the dark matter halo (while disturbed or
  asymmetric lobe morphologies indicate more dynamic/irregular gas distributions)
\item low-frequency lobe emission that extends a substantial distance
  from the hotspots back towards the nucleus, so that the internal
  pressure can be estimated reliably without hotspot contamination.
\end{itemize}
By applying these criteria, we expect to select radio galaxies for
which the magnetic field strength relative to equipartition and lobe
overpressuring at the midpoint are similar.

The FRII sample of \citetalias{ineson17} meets the first criterion listed above. We applied the second and third criteria to this sample, initially using our high-resolution radio maps described in \citet{ineson13} and \citet{ineson15}. The parent sample of \citetalias{ineson17} includes 33 radio galaxies with measured external pressure distributions, in two redshift ranges, $z<0.2$ and $0.4<z<0.6$. Applying the selection criteria above led to the exclusion of 6 objects: 3C\,303, 3C\,321 3C\,433, 4C\,73.08, 6C\,0850$+$3747, and 7C\,0219$+$3423. The remaining sources all have regular, symmetric ``classical double'' morphology and extended lobes, without significant signs of disturbance. Therefore, in this redshift range at least $\sim 80$ per cent of FRIIs are good candidates for our
method. Fig.~\ref{fig:exclude} shows examples of sources excluded due to being dominated by hotspot emission or having an irregular morphology. In the following sections we make use of our morphologically regular subsample of 27 objects, which spans three orders of magnitude in radio luminosity from $10^{25}$ W Hz$^{-1}$ sr$^{-1} < L_{\rm 151MHz} < 10^{28}$ W Hz$^{-1}$
sr$^{-1}$ and contains objects with linear sizes between 50 -- 1000 kpc.  While the ratio of regular to asymmetric/disturbed FRIIs may
evolve with redshift, we nevertheless expect to be able to construct very large samples of objects meeting our criteria from radio surveys.
 In Section~\ref{sec:disc} we comment on how well our morphological selection criteria can be applied to low-resolution radio surveys. We note that we make no selection based on accretion mode (low- or high-excitation class), as \citetalias{ineson17} have shown that there is no difference in the lobe composition or dynamics of FRII LERGs and HERGs.
 
\begin{figure*}
\centerline{\hbox{
\includegraphics[height=7.5cm]{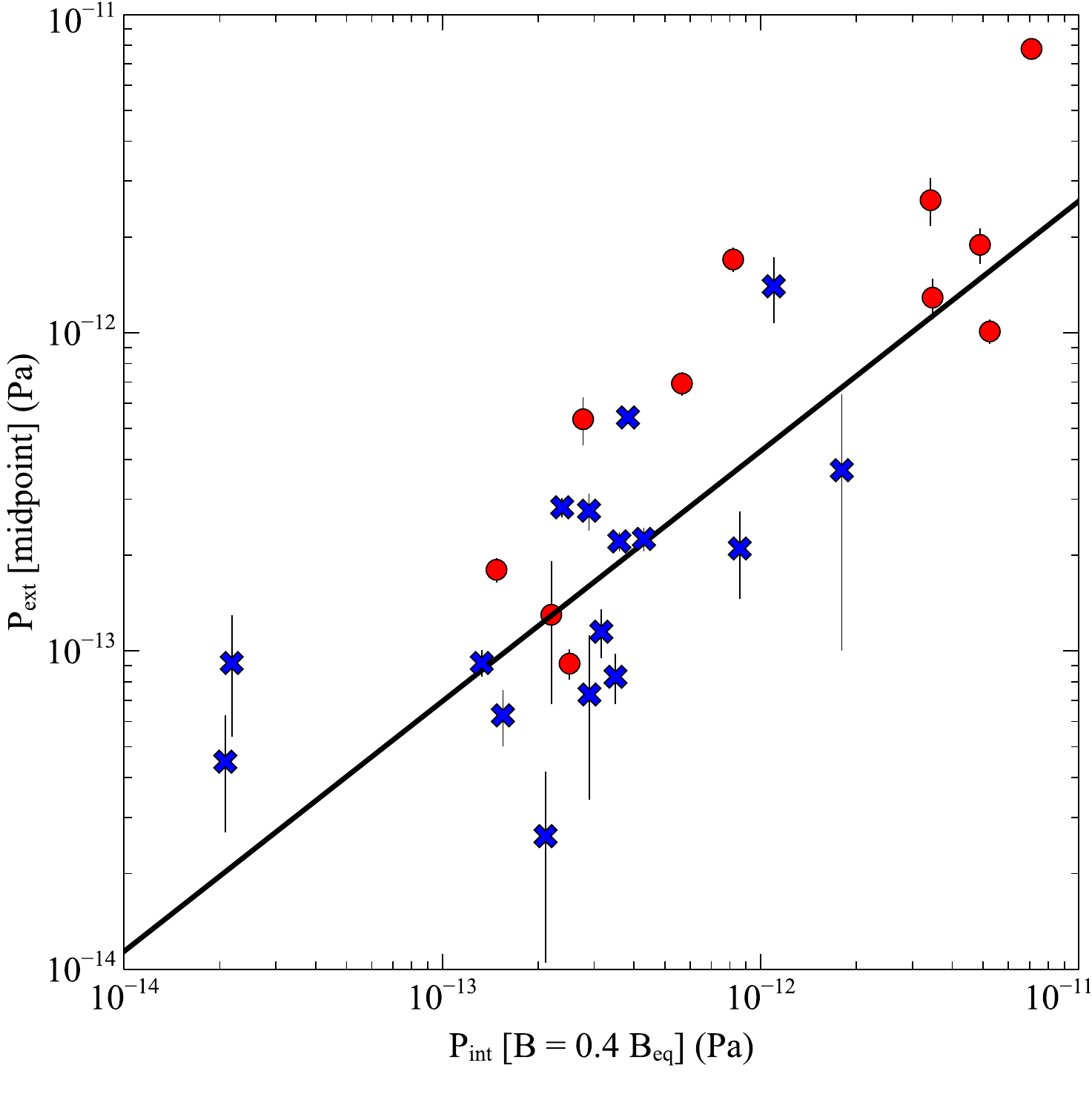}
\includegraphics[height=8.2cm]{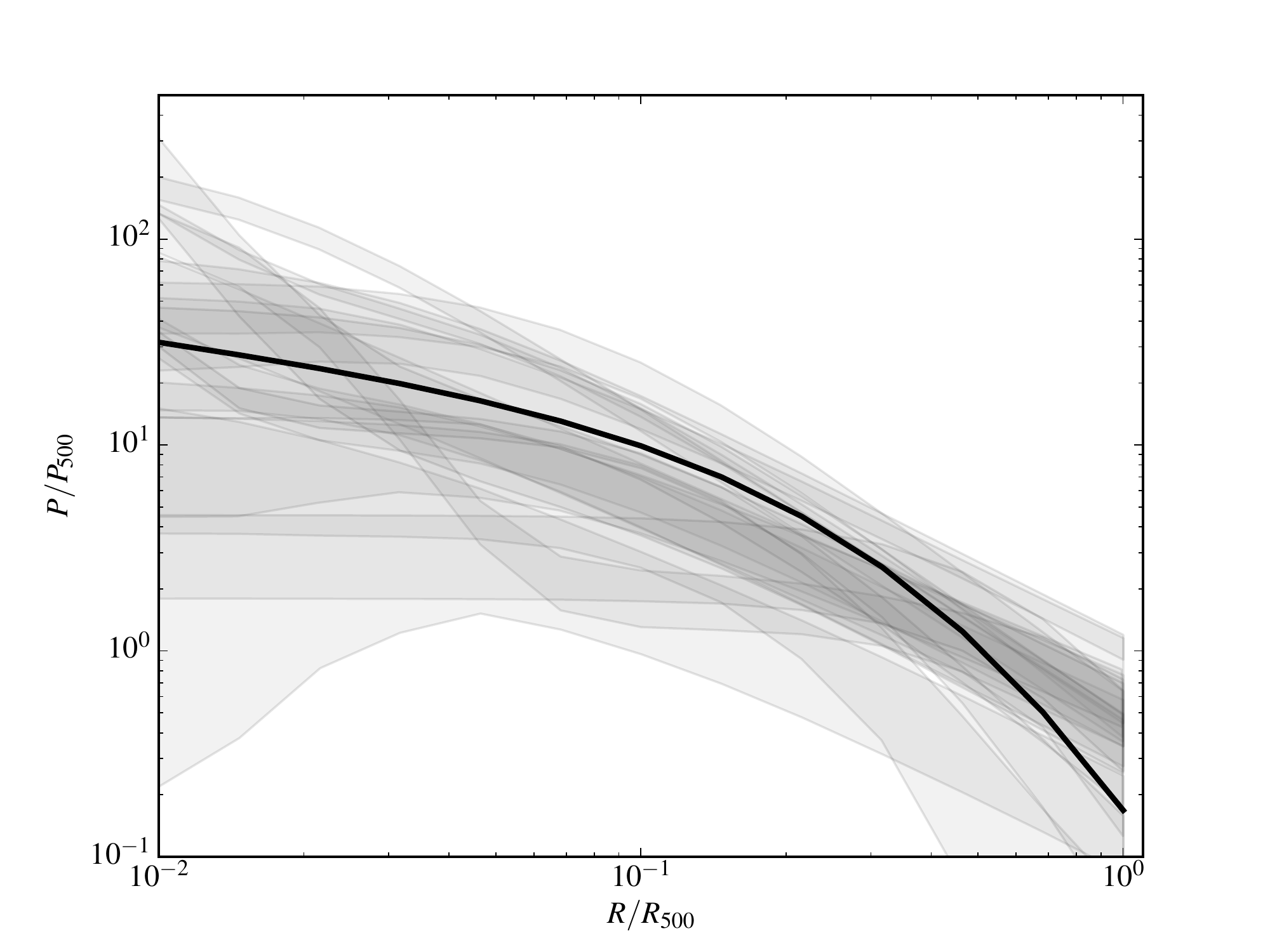}}}
\caption{Left: The internal lobe pressures calculated assuming $B=0.4B_{\rm eq}$
  vs. the measured external ICM pressure at the mid-lobe distance, for
  the morphologically selected subsample of \citetalias{ineson17}, demonstrating that
  this {\it radio-only} pressure estimate can reliably predict the
  external mid-lobe pressure (solid line is the best-fitting relation
  as given by Eq.~\ref{eq:peq}). Blue crosses indicate the low-$z$ sample
  ($z<0.2$) and red plus symbols the high-z sample ($0.4<z<0.6$), showing that the same relation holds for both subsamples, which implies that the relation is not
  driven by a distance dependence. Right: External pressure profiles for the subsample of \citetalias{ineson17} environments with well-constrained temperatures indicated by grey shaded areas showing 1$\sigma$ credible intervals. The universal profile of \citet{arnaud10} is plotted as a solid black line.}
\label{fig:pcomp}
\end{figure*}

\subsection{Predicting the external pressures around FRII radio lobes}
\label{sec:press}
We used the sample of 27 FRII radio galaxies obtained from the
selection of Section~\ref{sec:select} to investigate how well the external
ICM pressure could be predicted from the radio measurements alone. Our
assumption is that because the distribution of magnetic field
strengths for the FRII population is narrow, and the contribution of
protons is minimal (\citetalias{ineson17}, particularly figures 1 and 3 of that work), we can obtain a reasonable estimate of the internal pressure by assuming a magnetic field strength of $B = 0.4B_{\rm eq}$, which is the median value found in the IC analysis of \citetalias{ineson17}. 

For each radio galaxy we used the measured source geometry and radio
flux measurements from \citetalias{ineson17} to model the internal pressure on the
assumption of $B=0.4B_{\rm eq}$ and no protons. We used the {\sc synch}
code of \citet{hardcastle98} to determine the lobe internal pressures,
assuming a power law in electron energy between
$\gamma_{\rm min} = 10$ and $\gamma_{\rm max} = 10^{5}$, with an
energy index of $\delta=2.4$ (as discussed by \citetalias{ineson17}). This method can be
applied to large samples obtained from low-frequency surveys, with no need for additional information from other wavelengths other than the source redshift, provided that the source geometry and lobe radio flux can be measured reliably.

We compared our radio-estimated internal pressures ($P_{\rm int}$) with the
measured external (ICM) pressure for each source at the midpoint of
the modelled radio lobe ($P_{\rm ext}$), as shown in Fig.~\ref{fig:pcomp}. We find that our radio-estimated internal lobe pressures
are well correlated with the external mid-lobe pressures, consistent
with the finding of \citetalias{ineson17} that the lobes are typically close to pressure
balance at their midpoints. We used an MCMC least squares fitting code written using the {\sc emcee} package \citep{foreman13} to determine the best-fitting relation between our estimated values of $P_{\rm
  int}$ and measured $P_{\rm ext}$. Our best-fitting relation, with 90 per cent confidence intervals, is:
\begin{equation}
\label{eq:peq}
\log_{10}(P_{\rm ext}) = (0.79^{+0.02}_{-0.03})\log_{10}(P_{\rm int}) - (2.9\pm0.3)
\end{equation}
We find a slope a little lower than unity, which is consistent with a
pressure ratio that has some dependence on the internal pressure. This
is physically plausible as sources at high pressure may be expected to
be more overpressured even at the midpoint. We also carried out the
same comparison using the external pressure further out at the lobe
tip instead of the midpoint; however, we found that the internal pressure is a much weaker predictor of the lobe-tip pressure, which is as expected since \citetalias{ineson17} found a much wider range of external-to-internal pressure ratios at the lobe tip. 

Fig.~\ref{fig:pcomp} shows that there is no strong evolution of the best-fitting pressure relation with redshift: a Mann-Whitney U test of the distributions about this relation does not rule out at 95 per cent confidence the null hypothesis that the two redshift subsamples have the same parent population. We do, however, acknowledge that weak evolution of the relation of Eq.~\ref{eq:peq} cannot be ruled out by our data, and this should be investigated with deep X-ray surveys data and/or further deep X-ray observations of carefully selected samples.

\begin{figure*}
  \centerline{\hbox{
\includegraphics[width=5.6cm]{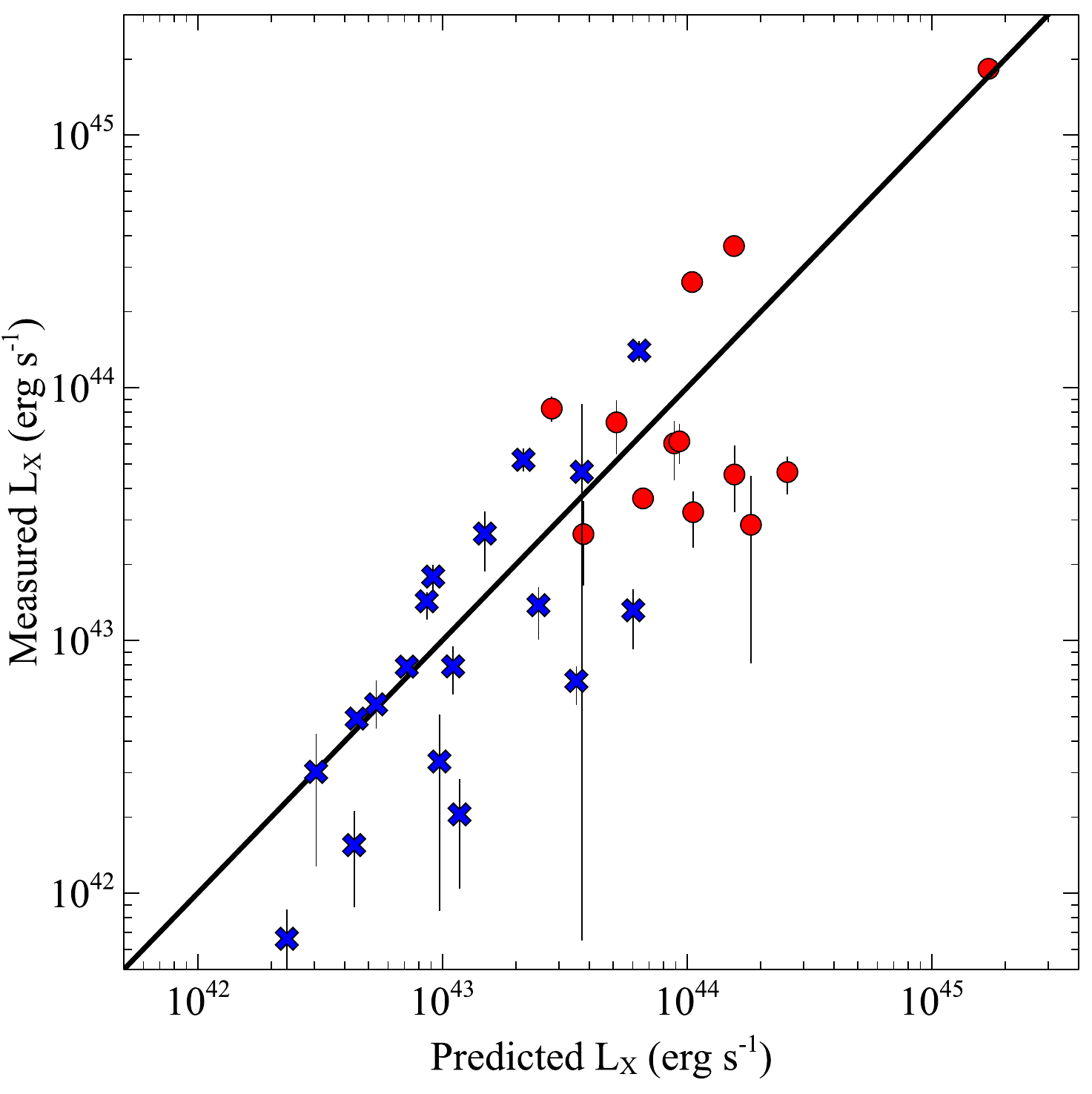}
\includegraphics[width=5.6cm]{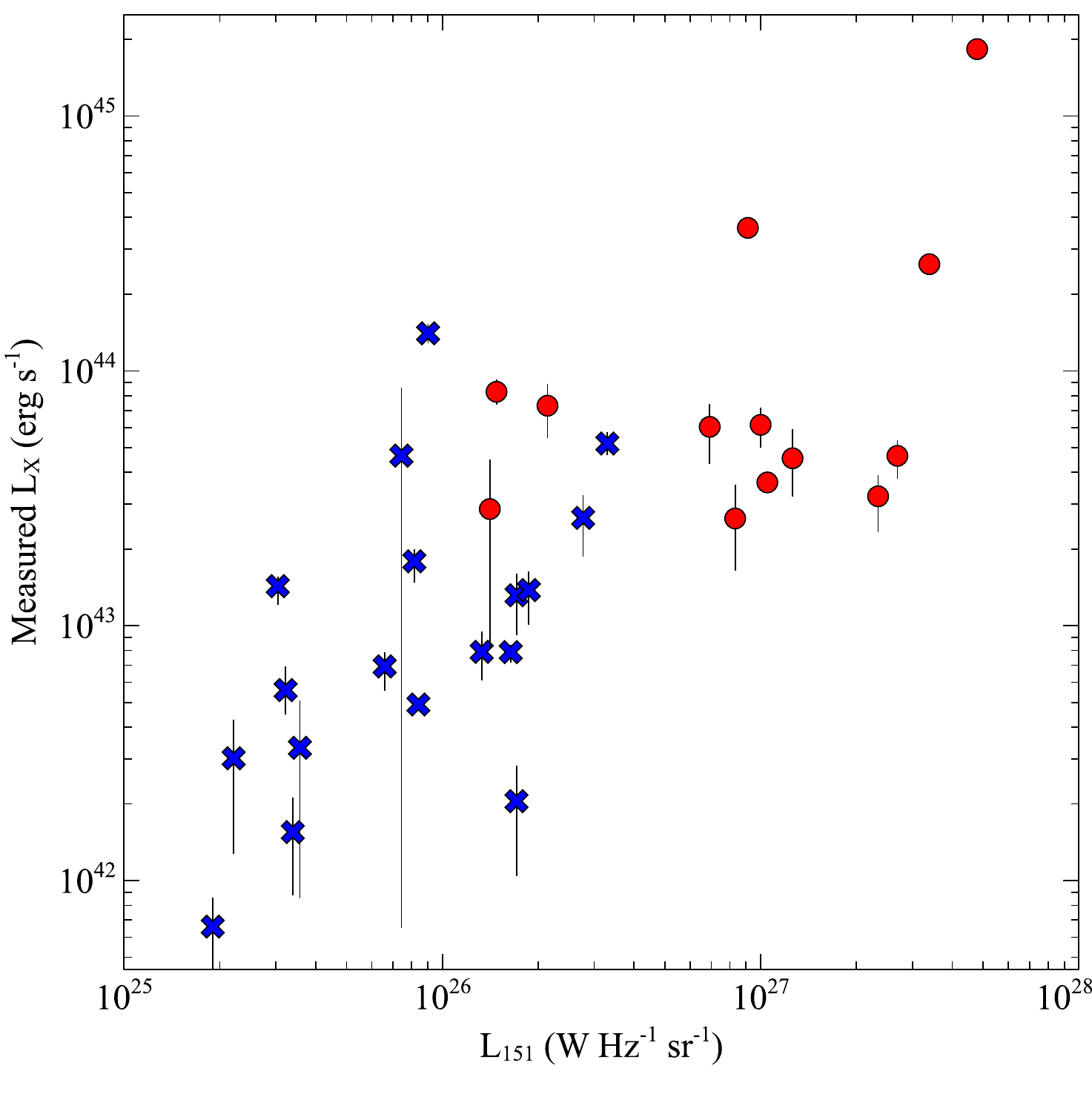}
\includegraphics[width=5.6cm]{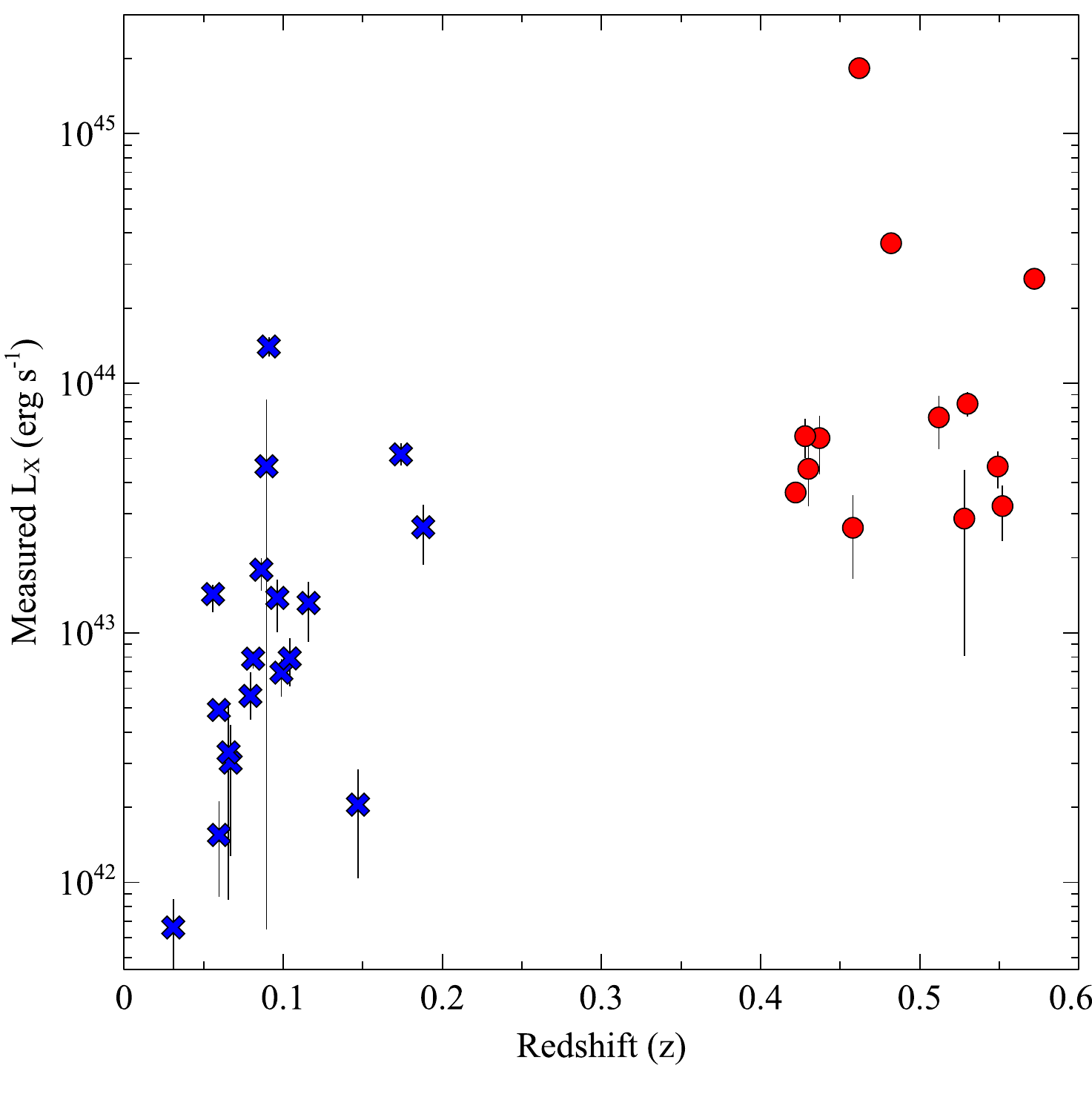}}}
\caption{Left: The predicted group/cluster X-ray luminosities
    determined from the radio-measured pressures for the
    morphologically selected subsample of \citetalias{ineson17}, obtained using the
    method described in Section~\ref{sec:envs}, compared with their measured X-ray luminosities, demonstrating that for this sample the environments
    can reliably be predicted from the (low-frequency) radio-estimated
    lobe pressures. A line of unity (not best fit) is shown as a solid
    line. Colours and symbols are as for the previous figure, showing that the
    prediction method is successful for both redshift ranges. Middle:
    The relationship between measured X-ray luminosity and radio
    luminosity, and right: the relationship between measured X-ray
    luminosity and redshift.}
    \label{fig:result}
\end{figure*}

\subsection{Using pressure to predict X-ray luminosity}
\label{sec:envs}
We have demonstrated that we can reliably predict the ICM pressure at
a single distance within the group/cluster surrounding an FRII radio
galaxy. To be able to predict the overall group/cluster richness, we
need to be able to translate this pressure measurement into an estimate of cluster mass and/or X-ray luminosity. We
therefore examined whether the universal cluster pressure profile of
\citealt{arnaud10} (hereafter \citetalias{arnaud10}) could be applied to FRII
environments. We first examined the subsample of \citetalias{ineson17} for which
reliable spectral temperature estimates were available. Our pressure
profiles have large uncertainties, and were obtained from either beta
or double-beta model fits to the surface-brightness profiles (see
\citetalias{ineson17}). The environments are also mainly galaxy groups
($L_{\rm X} < 10^{44}$ erg s${-1}$), and so lower in mass than the sample
of \citetalias{arnaud10}; however, \citet{sun11} have shown that the universal profile also applies on galaxy group scales. Fig.~\ref{fig:pcomp} (right) shows that our scaled
pressure profiles are in broad agreement with the \citetalias{arnaud10} profile shape
and normalization.

We therefore tested whether the universal cluster pressure profile
could be used to translate our single-radius pressure estimates, from
our radio predictions, into cluster mass and X-ray luminosity. We used
the profile of \citetalias{arnaud10} (Equations 11, 12 and 13 of that work) and our
input values of distance and $P_{\rm ext}$ to estimate $M_{\rm 500}$
for each radio-galaxy environment. We then applied the
$L_{\rm X}$--$M_{\rm 500}$ relation of \citet{pratt09} (using the
relation for $L_{\rm X} < R_{500}$) to obtain a predicted $L_{\rm X}$
for each radio-galaxy environment. Finally we compared our
radio-predicted X-ray luminosities with the measured X-ray
luminosities for our sample from \citet{ineson15}. Fig.~\ref{fig:result} (left) shows the remarkable agreement
between our radio-predicted X-ray luminosities and the measured values
across more than three orders of magnitude in X-ray luminosity (from
poor groups to moderate clusters) -- {\it our method appears to
  predict ICM X-ray luminosity to within roughly a factor of two based only on
  the measured properties of the embedded radio galaxy.} We have
included four additional FRII sources from the sample of \citetalias{ineson17} for which
external pressure distributions could not be measured, but which meet
the morphological selection criteria, leading to a sample of 31
objects, as listed in Table~\ref{tab:sample}. 

As shown in the middle panel of Fig.~\ref{fig:result}, there is a
correlation in our sample between the much simpler quantity of radio luminosity and
X-ray luminosity, which could also be used to predict FRII galaxy
group environments; however, this has much greater scatter (more than
an order of magnitude), and thus weaker predictive power. Our
sample does have a relationship between distance and measured X-ray
luminosity. We therefore show the low-$z$ ($z<0.2$) and high-$z$
($0.4<z<0.6$) subsamples of \citetalias{ineson17} in different colours in
Figs.~\ref{fig:pcomp} and~\ref{fig:result} to demonstrate that
the midpoint pressure relation and the predictive method both apply
separately to the two subsamples.  The right-hand panel of
Fig.~\ref{fig:result} also demonstrates clearly that there is no significant distance dependence of X-ray luminosity within each subsample, and hence the strong predictive relation of the left-hand panel cannot be driven by distance.

\section{Discussion}
\label{sec:disc}
Our method of predicting environmental X-ray luminosity based on radio
properties alone, presented in the previous section, appears extremely
successful; however, there are a number of questions that must be
addressed in order to be confident that the method will be useful for
large samples at higher redshift.

We must first consider how well our radio selection procedure can be applied to data from wide-field radio surveys. In the short-term, we hope to apply our method to the LOFAR Tier-1 surveys data \citep{shimwell17}, which has a spatial resolution of $\sim 6$ arcsec, corresponding to a physical resolution of $\sim 50$ kpc at $z=1$. We re-applied our morphological selection criteria using the TGSS Alternative Data Release (ADR) \citep{intema17} 150-MHz images for the \citetalias{ineson17} parent sample, which have a typical resolution of $\sim 25$ arcsec, so that the physical resolution at $z \sim 0.1$ is comparable to LOFAR Tier-1 resolution at $z \sim 1$. We found that 28/33 sources were sufficiently resolved to apply the criteria of \S~\ref{sec:select}, with all of the sources we originally excluded using the high-resolution data also being excluded using TGSS. A further five sources were excluded due to inadequate resolution, so that our method could have been applied to a sample of 23 objects had we only had data of TGSS resolution available. We therefore conclude that our morphological selection method can be applied accurately to LOFAR surveys out to $z>1$.

We used the LOFAR Bo\"{o}tes Tier-1 survey data presented by \citet{williams16} and \citet{williams17} to estimate the sky density of FRII radio galaxies at high redshifts. We take the assumption that all AGN above $L_{\rm 150MHz} = 3 \times 10^{26}$ W Hz$^{-1}$ are FRIIs, leading to estimated sky densities of $\sim 1.5$ FRIIs deg$^{2}$ between $0.5 < z < 1$, $\sim 2$ FRIIs deg$^{2}$ between $1 < z < 1.5$ and $\sim 1$ FRIIs deg$^{2}$ between $1.5 < z < 2$. The resolved fractions are 100, 90 and 82 per cent, respectively, for the three redshift bins (Williams, private communication). The adopted luminosity cut is conservative, as it is now known that sources of FRII morphology can be found at luminosities well below the traditional \citet{fanaroff74} divide \citep[e.g.][]{miraghaei17}, hence the true numbers could be higher. However, it is also known that source sizes in flux-limited samples evolve with redshift \citep[e.g.][]{blundell99} which could reduce the sample sizes that are classifiable with LOFAR data. Such size evolution is seen in a subset of the LOFAR Bo\"{o}tes sample examined by \citet{morabito17}, but their sample excluded low-excitation radio galaxies, and included only objects with firm spectroscopic redshifts, which may introduce host galaxy/environmental biases: the size evolution to $z = 2$ of the full LOFAR 150-MHz selected population is not yet well determined. If we assume, based on our TGSS comparison and consistent with the LOFAR--Bo\"{o}tes resolved fraction, that $\sim 2$/$3$ of the high-redshift sources will meet our morphological classification criteria, we would expect to be able to compile samples of $>5,000$ candidate galaxy groups at $z>1$, and $>2,000$ at $z>1.5$ with LOFAR Tier-1 surveys over the next few years, as the observed sky area accumulates\footnote{Around 2,500 deg$^{2}$ of mainly contiguous sky area has been observed to Tier-1 depth as of May 2017}. Size evolution may reduce these sample sizes by a factor of a few, but large numbers of >200-kpc sized objects should be detectable in this redshift range (e.g. $\sim 1$/$3$ of 3CRR objects at $z>1$ have linear sizes greater than 200 kpc). With the substantially higher resolution and sensitivity expected to be available with the SKA, it should be possible to extend the method to the redshifts where the earliest galaxy groups with a hot ICM are forming.

Of course another important concern is that our method relies on the applicability of the universal pressure profile of \citetalias{arnaud10} to FRII radio-galaxy environments over a large redshift range. We have demonstrated in Section~\ref{sec:press} that our sample
of radio-galaxy environments at $z<0.6$ have pressure profiles
broadly consistent with \citetalias{arnaud10}; however, our method will be most valuable if it can be applied to environments at $z>1$ where methods to identify and
characterize group-scale environments in particular are very
limited. A crucial question is therefore whether we would expect our method to apply at all redshifts. Since the radio lobes themselves are sensitive to
changes in the typical pressure distribution, if the typical ICM
environment for FRII radio galaxies evolves with redshift our method
should nevertheless accurately predict the external pressure at the
lobe midpoint. If the gas distributions of FRII environments evolve to
become typically more dynamic/irregular at $z>1$ then we would expect
that a larger fraction of the FRII population will fail to meet our
morphological selection criteria; however, those that do would be
expected to occupy relaxed gas distributions. We must then ask whether
the universal pressure profile of \citetalias{arnaud10} should continue to predict
$M_{\rm 500}$ correctly at $z>1$. As the formulae we use account for
standard evolution (e.g. of $\rho_{\rm crit}$) with redshift we would
expect the prediction of $M_{\rm 500}$ to be reliable. A more serious
concern is whether we would expect to see either a bias or increased
scatter in the relationship between $M_{\rm 500}$ and $L_{\rm X}$. The
relationship of \citet{pratt09} is for clusters at $z<0.2$; our
method includes standard self-similar evolution, but the evolution of
this relation at $z>1$ is not observationally tested. Any systematic
departure from this form of evolution, and particularly any increased
scatter in the relation between $L_{\rm X}$ and $M_{\rm 500}$ in
forming proto-groups and clusters will affect our method. 

While the caveats above show the need for some caution in applying our
method to high redshifts, it is important to emphasize that the radio
surveys that will provide the samples of high-redshift FRIIs for
environmental characterization should offer strong clues about the
effects of any environmental evolution: any systematic changes in the environments of the FRII radio galaxies will also be manifest as changes in the size, $L_{\rm R}$ and morphological distributions as a function of
redshift. It will be necessary to disentangle such effects from changes caused by other effects, such as evolution in AGN duty cycles; however, comparisons with hydrodynamical simulations of FRII radio galaxies in a range of environments \citep[e.g.][]{hardcastle14,english16} should enable careful diagnosis of the range of environments at $z>1$.

We therefore conclude that our method offers considerable potential
for the compilation of large samples of galaxy groups at $z>1$ with
LOFAR and out to $z>2$ with future surveys such as those planned with
the SKA. It will be essential to calibrate our method rigorously at moderate redshifts with X-ray observations of new samples of radio galaxies with unknown
large-scale environments. If this is successful then we will be able
to compile large galaxy group catalogues via embedded FRII radio
galaxies using LOFAR surveys data over the next few years.

\section{Summary}
We have described a new method by which embedded FRII radio galaxies
can be used to characterize group and cluster environments, which
offers the potential to compile large catalogues of evolved galaxy
groups at $z>1$. Our method exploits two important observational
results applying to the sub-population of `classical double' FRII
radio galaxies alone:
\begin{itemize}
\item the conclusion from X-ray inverse-Compton studies that FRII radio
  galaxies have a narrow distribution in magnetic field strength, with
  a median value of 0.4 $\times$ the equipartition field strength, and
  contain no energetically important protons.  
\item the conclusion from comparisons of external and internal
  pressures that FRII lobes are typically within a small factor of
  pressure balance at the lobe midpoint
\end{itemize}
These two results allow us to estimate reliably the internal
radio-lobe pressures of morphologically regular FRII radio galaxies
and therefore to estimate the external ICM pressure at the lobe
midpoint {\it from low-frequency radio measurements alone}. In particular, because our method is based on the lobe composition and dynamics, it applies to both low and high-excitation FRII radio galaxies.

We have tested our method by applying a simple morphological selection
to the FRII sample of \citetalias{ineson17}, demonstrating that the
X-ray luminosity of the host galaxy group or cluster can be predicted
to within a factor of $\sim 2$. It will be necessary to test
our method carefully using new samples of radio galaxies with unknown
environments, calibrating any evolutionary effects not accounted for by the current analysis; however, radio-lobe morphology is a sensitive probe of environmental gas distributions, and hence applying
careful morphological selection to high redshift samples is likely to
prove an effective means of finding, and characterizing the ICM of, large samples of galaxy groups and low-mass clusters. If such samples can be compiled they will be both of general value in the study of large-scale structure evolution and specifically useful for studies of radio-loud AGN feedback at high redshift.

\begin{table*}
\caption{Table of source properties for the \citetalias{ineson17} morphologically
  regular subsample. Fields are: source name, redshift, 151-MHz radio
  luminosity, lobe spherical radius (the radius of a sphere with
  equivalent volume to one lobe -- average for the two lobes), midpoint distance, estimated internal
  pressure (assuming $B=0.4B_{\rm eq}$), measured external pressure at
  the midpoint, and measured bolometric X-ray luminosity for
  $R<R_{500}$. All X-ray measurements and source geometries are taken from \citetalias{ineson17}; redshifts and radio luminosities are as compiled from the literature by \citet{ineson15}. Note that four
  sources, not included in Fig.~\ref{fig:pcomp}, do not have external pressure
  measurements, as explained in the text.}
\label{tab:sample}
\begin{tabular}{lrrrrrrr}
\hline
Source&$z$&$L_{151MHz}$&Lobe radius&Midpoint&$P_{\rm int}$&$P_{\rm ext,mid}$&$L_{X}$\\
&&$10^{25}$ W Hz sr$^{-1}$&kpc&kpc&$10^{-13}$ Pa&$10^{-13}$ Pa&$10^{43}$ erg s$^{-1}$\\
\hline
3C19&0.482&91.2&10.2&11.0&340&-&$36\pm3$\\
3C33&0.0596&8.4&59&86.4&2.1&$0.26^{+0.8}_{-0.23}$&$0.49^{+0.02}_{-0.03}$\\
3C35&0.067&2.2&185&239&2.1&$0.45^{+0.15}_{-0.21}$&$0.3^{+0.13}_{-0.17}$\\
3C46&0.437&69&167&293&2.8&$5.4^{+1.1}_{-0.7}$&$6.0^{+1.4}_{-1.7}$\\
3C98&0.309&1.9&41&42.6&2.9&$0.73^{+0.4}_{-0.5}$&$0.07\pm0.02$\\
3C192&0.0597&3.4&64&63&3.2&$1.2\pm0.2$&$0.16^{+0.06}_{-0.07}$\\
3C200&0.458&83.2&44&50&34&$26^{+0.6}_{-0.4}$&$2.6^{+0.9}_{-1.0}$\\
3C219&0.174&33&110&146&3.8&$5.4^{+0.2}_{-0.3}$&$5.2^{+0.6}_{-0.5}$\\
3C228&0.552&234&63&99&35&$13\pm2$&$3.2^{+0.7}_{-0.9}$\\
3C236&0.0989&6.6&433&662&0.1&-&$0.69\pm0.1$\\
3C244.1&0.43&126&44&97&48.9&$19^{+3}_{-2}$&$4.5\pm1$\\
3C274.1&0.422&105&197&275&2.5&$0.9\pm0.1$&$3.7\pm0.3$\\
3C285&0.0794&3.2&69&66&3.6&$2.2^{+0.1}_{-0.2}$&$0.56\pm0.1$\\
3C295&0.462&480&7.5&10.9&1890&-&$183\pm4$\\
3C326&0.0895&7.4&324&553&0.22&$0.92^{+0.45}_{-0.31}$&$4.7^{+4.0}_{-4.6}$\\
3C330&0.549&269&48&125&52.5&$10.1^{+0.9}_{-0.8}$&$4.6^{+0.7}_{-0.9}$\\
3C388&0.091&9.0&20.4&22.5&67.3&-&$14\pm1$\\
3C427.1&0.572&339&44&61&71&$78\pm4$&$26\pm3$\\
3C452&0.0811&16.4&95&106&2.4&$2.8\pm0.2$&$0.79^{+0.05}_{-0.07}$\\
3C457&0.428&100&232&367&1.5&$1.8\pm0.2$&$6.1\pm1.1$\\
6C1132$+$3439&0.512&21.4&90&176&5.7&$6.9\pm0.6$&$7\pm1$\\
6C1200$+3416$&0.53&14.8&58&111&8.2&$17^{+1}_{-2}$&$8.3\pm0.9$\\
6C0857$+$3945&0.528&21.9&157&387&2.2&$1.3^{+0.8}_{-0.4}$&$2.9^{+1.6}_{-2.1}$\\
PKS0038$+$09&0.188&28&57&57.4&10.8&$14^{+4}_{-3}$&$2.7^{+0.6}_{-0.8}$\\
PKS0043$-$42&0.116&17.1&31&96&18.4&$3.7\pm0.3$&$1.3^{+0.3}_{-0.4}$\\
PKS0213$-$13&0.147&17&58&58.1&8.6&$2.1^{+0.7}_{-0.6}$&$0.21^{+0.08}_{-0.1}$\\
PKS0349$-$27&0.0657&3.6&79&160&1.6&$0.63^{+0.2}_{-0.1}$&$0.3\pm0.2$\\
PKS0945$+$07&0.0863&8.2&64&108&2.9&$2.8^{+0.4}_{-0.3}$&$1.8^{+0.2}_{-0.3}$\\
PKS1559$+$02&0.1041&13.3&84&107&3.5&$0.8\pm0.1$&$0.8\pm0.2$\\
PKS2221$-$02&0.0559&3.0&98&163&1.3&$0.92^{+0.10}_{-0.08}$&$1.4^{+0.1}_{-0.2}$\\
PKS2356$-$61&0.0963&18.7&78&147&4.3&$2.3\pm0.2$&$1.4^{+0.3}_{-0.4}$\\
\hline
\end{tabular}
\end{table*}

\section*{Acknowledgments}

JHC, BM and MJH acknowledge support from the Science and Technology Facilities Council (STFC) under grants ST/M001326/1 and ST/M001008/1. We would like to thank Wendy Williams for providing unpublished information about AGN resolved fractions for the LOFAR-Bo\"{o}tes catalogue.




\bibliographystyle{mnras}
\bibliography{predict} 

\bsp	
\label{lastpage}
\end{document}